\documentstyle[eqsecnum,aps,preprint]{revtex} 
\tightenlines
\begin{document}
\draft
\preprint{IITK/Phy/96/}
\title{Geometrical representation of angular momentum coherence and 
squeezing}
\author{Abir Bandyopadhyay\footnote{ Present address : Centre for 
Theoretical Studies, Indian Institute of Science, Bangalore 560 012, INDIA.
Also at Jawaharlal Nehru Centre for Advanced Scientific Research, Bangalore 
560 064 (Do \underline{\bf NOT}  communicate to this JNC ASR address.)  
Electronic address :abir@cts.iisc.ernet.in} 
and 
Jagdish Rai\footnote{Electronic addresses :jrai@iitk.ac.in, 
jrai@rediffmail.com}}
\address{Department of Physics, Indian Institute of Technology, Kanpur -
208016, INDIA.}
\date{\today}
\maketitle
\begin{abstract}
        A simple, and elegant geometrical representation is developed to
describe the concept of coherence and squeezing for angular momentum
operators. Angular momentum squeezed states were obtained by applying
Bogoliubov transformation on the angular momentum coherent states in
Schwinger representation [Phys. Rev. {\bf{A 51}} (1995)]. We present the
geometrical phase space description of angular momentum coherent and
squeezed states and relate with the harmonic oscillator. The unique
feature of our geometric representation is the portraying of the
expectation values of the angular momentum components accompanied by their
uncertainties. The bosonic representation of the angular momentum coherent
and squeezed states is compared with the conventional one mentioning the
advantages of this representation of angular momentum in context of
coherence and squeezing.  Extension of our work on single mode squeezing
to double mode squeezing is presented and compared with the single mode
one.  We also point out the possible applications of the geometrical
representation in analyzing the accuracy of interferometers and in
studying the behavior, dynamics of an ensemble of quantum-mechanical
two-level systems and its interaction with radiation.
\end{abstract}
\section{Introduction}
     The harmonic oscillator phase space description of electromagnetic
fields has had great success in understanding the semiclassical and
quantum theories of coherent light \cite{klauder}. Coherent states are
defined generally as minimum uncertainty states. The product of the
uncertainties in the quadratures is minimum for these states and both the
quadratures have equal uncertainties. Squeezing redistributes the
uncertainties of the two quadratures resulting in one of them having less
than its value for the coherent state at the expense of increasing the
other \cite{sq}. The geometrical description of the Heisenberg uncertainty
relation of two noncommuting variables (quadratures) is well known to give
a better understanding of the inherent fluctuation due to the quantum
nature of the squeezed light. The quadratures of the phase space retain
the minimum uncertainty property of the coherent states.
    
    Angular momentum or SU(2) algebra describes the behavior of an
ensemble of two quantum-level noninteracting systems. Examples of these
systems include interferometers, non-interacting ensemble of two-level
atoms or molecules. Sensitivity of interferometers, when described by
SU(2) algebra, can be defined in terms of matrix elements of mean and
variances of angular momentum operators \cite{yurke}. Feynman {\it et al.}
\cite{feynman} have constructed a simple geometrical representation of the
Schrodinger equation for such two-level systems to solve maser problems
and radiation damping. In their approach the two-level quantum system is
described by the state vector $\vec r$ with components determined by the
probability amplitude. The dynamics of the system is described by the
differential equation of the state vector ${{d{\vec r}}\over {dt}}={{\vec
\omega}\times {\vec r}}$, where $\vec \omega$ is the angular velocity.
However, in their approach there is no description of the uncertainty
associated with the state vector. These uncertainties are important in
describing the full quantum nature of the system e.g., the interaction of
the system with quantized radiation. Our work is an extension of the
Feynman {\it{et al.}}'s description as we have included the uncertainties
associated with the state vectors. Considerable amount of work has also
been done in order to understand the coherent angular momentum states
\cite{nuref1,arecchi}. Arecchi had constructed the angular momentum
coherent states by rotating the angular momentum ground state
\cite{arecchi} similarly as of displacement operator coherent states for
electromagnetic waves. During last few years several authors have tried to
construct squeezed angular momentum states and study their properties in
terms of atomic systems \cite{nuref2,gsa,kita}.

    Schwinger developed an abstract algebra which perfectly describes
the angular momentum systems \cite{sch}. The algebra assumes the angular
momentum operators as a combination of two sets ($\pm$) of uncorrelated
creation and annihilation operators of spin $\pm {1\over 2}$ which obey
the bosonic commutation relations. Apart from the usual rotational angular
momentum states, this algebra also describes the pseudo-angular momentum
systems such as interferometry or a collection of two level systems proving
to be quite a success. Atkins and Dobson (AD) had constructed angular
momentum coherent (SAMC) states using Schwinger representation about one
and a half year before Arecchi \cite{ad}. They had connected these states
to two dimensional oscillator states quantized in orthogonal direction.
The matrix elements of the components can be calculated easily from the
expressions of their operator form. In the classical limit the SAMC states
become the classical vector $\vec J$ in the sense that they behave like
their classical analogs and their variances are small compared to the
absolute value of their averages. They have used these states to study
elliptically polarized states successfully \cite{ab}. These SAMC states
have also been used to describe the rotational levels of nuclear and
moleculer systems by Fonda {it{et al.}} \cite{fonda}.

    Recently, we have proposed a technique for the generation of angular
momentum squeezed states \cite{abir1}. By combining the Schwinger angular
momentum representation with boson operators and the concept of squeezing
of bosons via Bogoliubov transformation, we were successful in producing
states that exhibit squeezing of angular momentum operators. We have
called these states Schwinger Angular Momentum Squeezed (SAMS) states. Our
procedure was an extension of the work of Atkins and Dobson \cite{ad} on
angular momentum coherent (SAMC) states. We also found application of the
SAMS states in enhancing the sensitivity of interferometers and in study
of two level atoms \cite{abir1}. In this paper we have constructed an
elegant geometrical representation of the concept of coherence and
squeezing for the angular momentum operators. We present some new results
for two mode squeezing and discuss their geometrical behavior and
non-usefulness in this paper. The results of SAMC states and single mode
SAMS states are also considered for the geometrical phase space realization
of these states. Our representation, combined with the picture given by
Feynman {\it{et al.}}, ensures to be an important tool to study the
interaction between radiation and coherent or squeezed two-level
systems.

    The paper is organized as follows. In sec. II we recapitulate some
results of the Schwinger Angular Momentum Coherent (SAMC) states to
describe them geometrically and compare them with the bosonic counterpart.
In the next section (III) we present the new two mode squeezing results
(IIIb) along with relevant single mode SAMS state results (IIIa) to
understand the phase space geometry. Finally we conclude comparing our
technique with other works on coherent angular momentum in context of
geometrical understanding of the phase space. We also furnish the possible
applications of the geometrical representation in the conclusion.

\section{Angular Momentum Coherent States}
    Schwinger \cite{sch} developed the entire angular momentum algebra in
terms of two sets(up and down) of uncorrelated harmonic oscillator
creation and annihilation operators constructing the angular momentum
operators as 
\begin{mathletters}
\label{jdefn}
\begin{equation}
    {J_+}={J_x} + i{J_y} = {{a_+}^\dagger}{a_-}
\end{equation} 
\begin{equation}
    {J_-}={J_x} - i{J_y} ={{a_-}^\dagger}{a_+} 
\end{equation}
\begin{equation} 
    {J_z} = {1\over 2}({a_+}^\dagger {a_+}
    -{a_-}^\dagger {a_-}).
\end{equation}
\end{mathletters}
The operators ${a_\pm}^\dagger ({a_\pm})$ create (annihilate) a
$\pm{1\over 2}$ spin and follow the bosonic commutation relations which
means that the whole system is considered as a combination of two sets of
boson states. This construction satisfies the standard angular momentum
commutation relation $[J_l,J_m]=i\epsilon_{lmn} J_n$. The angular momentum
basis states $\vert j,m\rangle$ can thus be created by action of the 
oscillator operators on the vacuum spinor $\vert 0,0\rangle$ 
\begin{equation}
    \vert j,m\rangle = \vert j+m\rangle_+\otimes\vert j-m\rangle_- = 
[(j+m)!(j-m)!]^{-{1\over 2}}{({a_+}^\dagger)}^{j+m}{({a_-}^\dagger)}^{j-m}
\vert 0,0\rangle .
\end{equation}

    AD \cite{ad} constructed the Schwinger Angular Momentum Coherent
(SAMC) states as the simultaneous eigenstates of the operators $a_\pm$.
They have shown that the SAMC states are minimum uncertainty states for
the angular momentum-angle uncertainty relation in the large $N (N={n_+}+
{n_-}\geq 10)$ limit \cite{ad}. According to their definition the
angular momentum coherent states $\vert \tilde{\alpha}\rangle = \tilde{D}
\vert 0,0\rangle$, with $\tilde{D} = D_+ (\alpha_+ ) D_- (\alpha_- )$, obey,
\begin{equation}
 a_{\pm} \vert{\tilde{\alpha}}\rangle = a_{\pm}\vert\alpha_+,\alpha_-\rangle =
 \alpha_\pm\vert\alpha_+ ,\alpha_- \rangle
\end{equation}

    The expansion of $\vert\tilde{\alpha}\rangle$ in terms of angular momentum
basis is given by
\begin{equation}
    {\vert\tilde{\alpha}\rangle}={e^{-{1\over 2}N}{\sum_j}{\sum_m}
{{(2j!)}^{-{1\over2}}}{{{2j} \choose {j+m}}^{1\over
    2}}{{\alpha_+}^{j+m}}{{\alpha_-}^{j-m}}{\vert j,m\rangle}}
\end{equation}

    The $\infty$:1 mapping of $\tilde{\alpha}$ onto
${(\langle J_x\rangle ,}{\langle J_y\rangle,}{\langle J_z\rangle )}$ is a 
consequence of the 2:1 homomorphism of
SU(2) and SO(3) \cite{ref14}. SU(2) is spanned by the subset of spinors of
length $\sqrt{N}$ and SO(3) is spanned by the subset of vectors of length
$\langle J\rangle$. The spinor $\vert\tilde{\alpha}\rangle$ is a vector 
sum of different
vectors $\vert j,m\rangle$ in the physical angular momentum space. Once the four
parameters in $\alpha _\pm$ are fixed, it automatically fixes the values
of $j$ and $m$ in the angular momentum space. To calculate the mean and
variances of the angular momentum components one has to use the
expressions of them in SAMC basis and fix the corresponding parameters.
The matrix elements can be expressed in terms of any set of these
parameters (${\alpha _\pm}$ or $j$ and $m$). The expressions in terms of
the angular momentum parameters give better understanding in the physical
space. For this reason we have expressed the matrix elements in terms of
angular momentum parameters throughout the paper.

    The mean of the angular momentum components are calculated as
following
\begin{mathletters}
\begin{equation}
\langle {J_x}\rangle ={\sqrt {j^2 -m^2}}\cos \Theta
\end{equation}
\begin{equation}
\langle {J_y}\rangle ={\sqrt {j^2 -m^2}}\sin \Theta
\end{equation}
\begin{equation}
l\angle {J_z}\rangle =m
\end{equation}
\end{mathletters}
where $\Theta = ({\theta _+}-{\theta _-})$ and the variances are
\begin{equation}
\label{eqn3.8b} 
{{\Delta J_x}^2} = {{\Delta J_y}^2} = {{\Delta J_y}^2} = {1\over 2}j
\end{equation}

    The above results show that the average value of $\vec {J}$ {\it
{i.e.}} the tip of it lies on a sphere of radius $j$. This can easily
be verified by squaring and adding the mean values of the angular momentum
components. The fluctuations of the components are also same in the three
directions. The equation of the region of uncertainty creates a sphere of
radius ${\sqrt{{3j}\over 2}}$ about the tip of the vector. This is shown
in Fig.~1. The polar angle of the state vector in the three
dimensional phase space is actually realized to be $\Theta$, the
difference between two phase angles of the coherence parameters.

     We have shown the uncertainties of harmonic oscillator and a schematic 
of In three dimensional phase space we have shown the uncertainty sphere
  of the angular momentum vector in Fig.~1(c). This is also
compared with the uncertainty circle of the harmonic oscillator in two
dimensional phase space in Fig.~1(a). For SAMC
states the radii of these uncertainty spheres (= $\sqrt{{3j}\over 2}$)
depend only on the radii of the mean sphere (= $j$) on which the tip of
the vector lies. For a fixed $j$ value the radii of the uncertainty
spheres does not vary on its position on the sphere for the choice of the
parameters $m$ or $\Theta$.  The uncertainty spheres has a circular
projection (uncertainty circle) in the X-Y plane. We have shown some 
uncertainty spheres on it for some different positions
(different values of $m$ and $\Theta$) in Fig.~1(c).  The
positions for $\vert m\vert \langle j$ can be compared with the displacement of
the ground state in the phase space picture for harmonic oscillator.  The
displacement of the uncertainty circle in the phase space of harmonic
oscillator is performed by the rotation of the uncertainty sphere in
corresponding three dimensional phase space for SAMC states. The position
after rotation is governed by the values of the parameters $m$ and
$({\theta _+}-{\theta _-})$.

    The uncertainty relation corresponding to the commutation relation
between the components of the angular momentum, $[J_l,J_m]=i\epsilon_{lmn}
J_n$, is
\begin{equation}
    {{\Delta J_x}^2}{{\Delta J_y}^2} \geq {{1 \over 4}{\vert{\langle 
{J_z}\rangle}\vert}^2}
\end{equation}
Putting the expressions of the matrix elements in the last equation one
can check that the equality occurs at $m=\pm j$. The two solutions for the
equality show the SU(2) symmetry of the system. Other values of the
angular momentum projection have uncertainties of both the quadratures
equal and same as the extremum cases. Though the non-extremum cases does
not violate the uncertainty relation in the last equation but the relation
is an equality only for extremum cases. Anyway, for physical purposes we
are interested in the absolute uncertainties in the quadratures which
remain same for all the SAMC states with same $j$ value.

\section{Angular Momentum Squeezed States}
\label{sec3}
    Following the work of AD on angular momentum coherent states, we have
generated squeezed angular momentum states by operating the squeezing
operators of the bosonic states on the SAMC states \cite{abir1}. For the
two mode ($\pm$) bosonic case the squeezing operators can be defined as
\begin{equation}
    S_\pm (\xi_\pm )=exp \left[{1\over 2}(\xi_\pm {{a_\pm}^{\dagger}}^2
    -{\xi_\pm}^\ast {{a_\pm}^2})\right] ,\quad {\xi_\pm} =
{r_\pm}{e^{i{\phi_\pm}}} 
\end{equation}
    We have created the general SAMS states by operating the two mode
squeezing operator $\tilde{S} = S_+ ( \xi_+ ) S_- ( \xi_- )$ on the SAMC
states
\begin{equation}
    \vert\psi\rangle ={S_+}({\xi_+}){S_-}({\xi_-}){D_+}({\gamma_+}){D_-}
({\gamma_-}) {\vert0,0\rangle}
\end{equation}
    For convenience in calculation we use the relation to interchange the
order of the squeezing and displacement operators \cite{fisher} and write
the general SAMS states as
\begin{equation}
\vert\psi\rangle ={D_+}({\alpha_+}){D_-}({\alpha_-}){S_+}({\xi_+}){S_-}({\xi_-})
    {\vert 0,0\rangle} 
\end{equation}
where $\gamma_\pm$ = $\cosh{r_\pm}{\alpha_\pm} + {e^{i{\phi_\pm}}}
\sinh{r_\pm}{{\alpha^\ast}_\pm}$.

\subsection{Single Mode Squeezing}
    The calculation of the expectation values and variances of the
operators of our interest is cumbersome due to the dependence on the large
number (eight) of parameters involved in the general SAMS states. First we
consider the squeezing in one mode only. At this point we do this for
simplicity though the utility of this choice will be clear in the
following subsection. The SU(2) symmetry tells us that we can choose any
one of the modes for squeezing.  So we choose to squeeze in the + mode
which reduces the expression of the basis state vectors of single mode SAMS 
states to
\begin{equation}
    \vert {\Psi}\rangle = \tilde{D} {S}({\xi})\vert 0,0\rangle
\end{equation}
where we have dropped the unnecessary suffix +.

    Calculating the mean of the angular momentum components \cite{abir1}
\begin{mathletters}
\label{eqn4.5}
\begin{equation}
    \langle{J_x}\rangle ={\sqrt {j^2 -m^2}}\cos {\Theta }
\end{equation}
\begin{equation}
    \langle{J_y}\rangle ={\sqrt {j^2 -m^2}}\sin {\Theta }
\end{equation}
\begin{equation}
    \langle{J_z}\rangle = m+{1\over 2}{{\sinh ^2}r}
\end{equation}
\end{mathletters}
we see that the traversing spherical surface of the mean of the tip of the
angular momentum vector has been changed to a prolate ellipsoid with same
axes in X and Y. The expression for the mean value of the angular momentum
projection or the Z-axis of the ellipsoid show an increase as squeezing is
increased. This is shown in Fig.~1(d).  It is to be noted that squeezing
in the other mode will change the mean sphere to an oblate ellipsoid with
$\langle{J_z}\rangle = m-{1\over 2}{{\sinh ^2}{r_-}}$ instead of a prolate one.
However, the other two axes will not change due to the choice of mode of
squeezing. We calculated the variances to get a feel of the uncertainty
nature of the SAMS states as
\begin{mathletters}
\begin{equation}
    {\Delta {J_x}^2} = {1\over 2}j + {1\over 2}{\sinh r}\bigl[{1\over
    2}{\sinh r} \{1+2(j-m)\} + (j-m){\cosh r}{\cos\delta}\bigr]
\end{equation}
\begin{equation}
   {\Delta{{J_y}^2} }= {1\over 2}j + {1\over 2}{\sinh r}\bigl[{1\over
    2}{\sinh r} {\{1+2(j-m)\} - (j-m)}{\cosh r}{\cos\delta}\bigr]
\end{equation} 
\begin{equation}
    {\Delta{J_z}^2} = {{j+m}\over 4}[e^{2r}{\cos^2 \eta}+e^{-2r}{\sin^2
    \eta}]+{1\over 2}{\sinh^2 r} {\cosh^2 r}+{{j-m}\over 4}
\end{equation}
\end{mathletters}
where $\delta =2{\theta_-} - \phi$ and $\eta ={\theta_+}-{\phi\over 2}$.
$r$ and $\phi$ are the squeezing parameters of the + mode. The effect of
squeezing in the other mode on the variances of the angular momentum
components can be obtained from the last equation by interchanging the
suffixes which is a consequence of the SU(2) symmetry. The uncertainty
ellipsoid for $m =-j$ have been shown in Fig.~1(d). It is to be noted that
all the axes of this uncertainty ellipsoid are different. This results to
an ellipsoidal projection on the X-Y plane. The squeezing of angular
momentum can be compared with the harmonic oscillator squeezing, which is
shown in Fig.1(b).

    In Fig.~2 we have plotted the dependence of the axes of the projected
uncertainty ellipses on the squeezing parameter $r$. The projected
uncertainty circle on the X-Y plane for the SAMC states are transformed to
ellipses, but with greater area (uncertainty product). It is clear from
the expressions that the maximum squeezing {\it{i.e.}} minimum fluctuation
of the squeezed quadrature occurs to the minimum uncertainty circle at
$m=\pm j$ as expected physically. From Fig.~2 it is clear that the minimum
uncertainty circle is squeezed (length of the semiminor axis is reduced)
up to a critical value of r (=$r_{min}$) though its area (uncertainty
product) is increased throughout. After that critical value of r the
length of both the axes of the uncertainty ellipse increases.

    It is interesting to note that squeezing in the + or - mode results
squeezing of uncertainty in $J_y$ and $J_x$ respectively. This means that
the squeezing in the angular momentum quadratures are directly related to
the mode of squeezing. It will be interesting to express the squeezing
operators in terms of operators in X-Y coordinates instead of $\pm$ to
identify the reason and exact mapping between them.

\subsection{Double Mode Squeezing}

    Now we consider the case of double mode squeezing. In the last
subsection we have squeezed only in one mode for the sake of simplicity
and promised to give the practical reasoning for this simplification in
this subsection. Actually two mode squeezing does not help in reducing the
uncertainty of any of the quadratures which we will show now. We can claim
from the results of the last subsection that if we squeeze both the modes
the uncertainties of the quadratures will be squeezed and expanded
simultaneously. The squeezing of the second mode in effect reduce the
amount of squeezing achieved by the first mode squeezing.

    We have calculated the expectation values of the components of the
angular momentum for double mode squeezing as
\begin{mathletters}
\begin{equation}
  \langle{J_x}\rangle = {\sqrt{{j^2}-{m^2}}}\cos\Theta
\end{equation}
\begin{equation}
  \langle{J_y}\rangle = {\sqrt{{j^2}-{m^2}}} \sin\Theta
\end{equation}
\begin{equation}
  \langle{J_z}\rangle = m + {1\over 2}({\sinh ^2}{r_+} - {\sinh ^2}{r_-}).
\end{equation}
\end{mathletters}
The expectations of the angular momentum components in X and Y direction
are seen to be same as that of SAMC states with no effect of squeezing.
The mean sphere is clearly seen to be transformed to an ellipsoid in
general with same X and Y axes. The Z axis of the ellipsoid will increase
or decrease as difference of squares of the hyperbolic sine functions of
the two parameters $r_\pm$. This affects the shape of the mean spheroid to
prolate or oblate. However, the expectation value of $J_z$ can be made to
be same as that of SAMC states by squeezing both the modes equally. This
will make the mean ellipsoid to be same mean sphere as for SAMC states.

    To show that the effect of double mode squeezing does not help in
squeezing of the angular momentum quadratures we have calculated the
uncertainties for some special choice of parameters. We have chosen the
phases in the squeezing parameters to be equal to zero and the magnitudes
of the squeezing parameters to be equal to $r$. This choice does not
affect the basic motivation of representing the states geometrically or
prove the disadvantage of double mode squeezing. We have calculated the
uncertainties in the angular momentum components for this special 
\cite{coment} choice as
\begin{mathletters}
\begin{eqnarray}
  \Delta{{J_x}^2}  &=&  {1\over 2}[j + \sinh {r}\{\sinh {r} + \cosh
  {r}\cos (2{\theta _+})\}(j+m) \nonumber \\
   &&   + \sinh {r}\{\sinh {r} + \cosh {r}~\cos (2{\theta _-})\}(j-m)
 + {\sinh ^2}{r}(1+\cosh 2r)]
\end{eqnarray}
\begin{eqnarray}
  \Delta{{J_y}^2} & =& {1\over 2}[j+ \sinh {r}\{\sinh {r} - \cosh~r~ 
\cos (2{\theta _+})\}(j+m) \nonumber \\
   && + \sinh {r}\{\sinh {r} - \cosh {r} \cos (2{\theta _-})\}(j-m) 
 + {\sinh ^2}{r}(1+ \cosh 2r)]
\end{eqnarray}
\begin{eqnarray}
  {\Delta{J_z}^2} & =& {{j+m}\over 4}[e^{2{r}}{\cos^2 {\theta _+}} + 
  e^{-2{r}}{\sin^2 {\theta _+}}] +{{j-m}\over 4}[e^{2{r}}{\cos^2 {\theta _-}} 
  + e^{-2{r}}{\sin^2 {\theta _-}}] \nonumber \\
  &&+{\sinh^2 {r}} ~~{\cosh^2 {r}} 
\end{eqnarray}
\end{mathletters}
    The uncertainty ellipsoids on the mean ellipsoid are similar as
single mode squeezing case. The uncertainties of the quadratures and their
product are plotted in Fig.~3 for $j$=50, $m$=-50, and $\theta_+ =
\theta _- =0$. Here we have plotted the results for the extremum
projection states which drops out the first term in both the quadrature
uncertainties. The squeezing in the uncertainty of $J_y$ prove our claim
that double mode squeezing deteriorates the effect of single mode squeezing
which was expected from qualitative reasoning. Squeezing both the modes by
same amount retains the SU(2) symmetry of the system but the choice of the
phases and magnitudes of coherent parameters breaks it resulting different
expressions and curves for the quadratures. This choice has been made to
show the difference distinctly and the dependence on the coherence phases.
With all the parameters same for the two modes one can show that the two
quadratures will behave similarly. We do not present the geometrical
pictures for the double mode squeezing as they are similar to the single
mode squeezing.

\section{Conclusion} 
    We have represented the angular momentum coherent (SAMC) and squeezed
(SAMS) states geometrically and studied their properties for a simple
choice of parameters. These simplifications does not hamper the
qualitative geometrical interpretation of these states. Actually,
consideration of all the parameters makes the results complicated and not
easily visible in the phase space picture. Due to this reason we have
simplified the results by these choices. We have also shown that two mode
squeezing deteriorates the squeezing effect in angular momentum
quadratures. We applied the SAMS states in analyzing the sensitivity of
interferometry \cite{abir1}. The effect of two mode squeezing on
interferometry can be seen from the results of the uncertainties in two
mode squeezing. As two mode squeezing increases the uncertainty of the
squeezed quadrature it will also increase the value of minimum detectable
phase difference ( $\Delta\Phi$ ) of any interferometer using beam
splitters. The relation between them in the frame rotated by $\pi\over 2$
about X axis can be written as ${\Delta\Phi} = {{\Delta J_y}\over {\vert
\langle J_x\rangle\vert}}$ \cite{abir1}.

  Any ensemble of two quantum-level system ({\it e.g.} atoms or molecules
can be considered as a spin-$1\over 2$ particle is described by the SU(2)
algebra of angular momentum systems \cite{nuref1,arecchi}. The number
operators (${\hat n}_\pm = j\pm m$) in the case of interferometric
representation correspond to the population or occupation numbers in the
upper and lower states of the system. In fact the complete set to describe
the system is achieved by adding a permutation group $P_N$ to the SO(3)
group. However, this does not change the basic essence of the formalism.
Squeezed atomic states were constructed by preferential population
distributions \cite{abir1}. Feynman {\it{et al.}} \cite{feynman} have
shown that the components of the pseudo angular-momentum vector completely
specifies the state of the system semiclassically. The power of the
geometrical method developed by Feynman {\it et al.}, lies in visualising
and solving problems involving transition between two quantum levels. For
example, the two classic problems discussed by them, the beam type maser
oscillators and the radiation damping, could be visualised very clearly by
the orientation of the state vector. Later application has led to elegant
method for visualising and solving the photon echo problem also.
Geometrical methods are found useful for problems that can be solved
analytically. But it can also provide valuable insight into the behavior
of the prowesses that are insolvable by analytical technique. Fermion
interferometry can be considered by changing the bosonic commutation
relation to the anticommutation relation for fermions. This is a totally
new possibility as no fermion analog of the boson squeezed states has yet
been found \cite{ref17}.

    Arecchi \cite{arecchi} had developed the coherent atomic states by
rotating the minimum uncertainty Dicke state (lowest projection state
$\vert j,-j\rangle$) in three dimensional phase space. Recently, the authers in
Ref. \cite{gsa} have described the Dicke states as a cap and annular
surfaces. As these states are eigenstates of $J_z$ they should not be
described by cap or annular surfaces with nonzero $\Delta J_z$ on the mean
sphere of the angular momentum vector. Instead the different Dicke states
($\vert j,m\rangle$) should be represented by an uncertainty circle. The Arecchi
type atomic coherent states have some other geometrical representation
problem wcich is not present in our picture. Their coherent states are
actually some rotated angular momentum ground states. The projections of
these rotated states in X-Y plane is an ellipse and thus show squeezing.
Physically, mere rotation should not change the status of the state. This
question has been raised by Kitagawa {\it {et.al}} \cite{kita} that if
these Arecchi type coherent states describe squeezed angular momentum
states under suitable choice of coordinates. Moreover, the area of the
projected ellipse, which is a measure of the uncertainty product, is
reduced from the area of the uncertainty circle of the ground state. This
is a direct violation of Heisenberg uncertainty principle. In our
definition of angular momentum coherent or squeezed states we have
overcome this difficulty in representation and answered the question of
Kitagawa {\it{et al.}}. The geometrical picture developed by us for SAMC
or SAMS states does not have this ambiguity and thus is a better
representation for the angular momentum coherent and squeezed states.

 
\begin{figure}
\vskip 14.2cm
\includegraphics{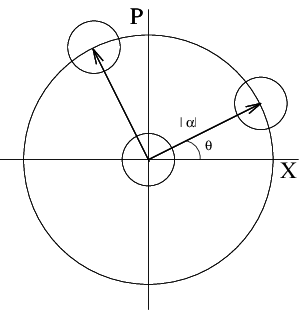}
\includegraphics{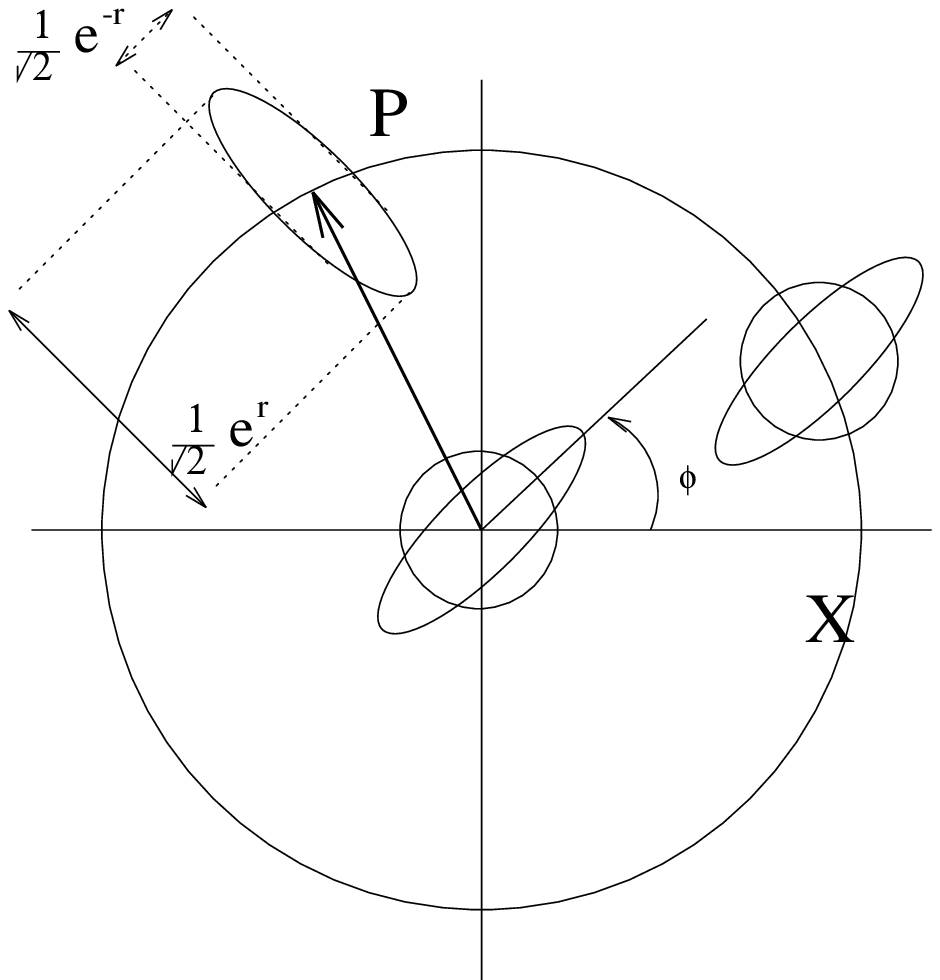}
\includegraphics{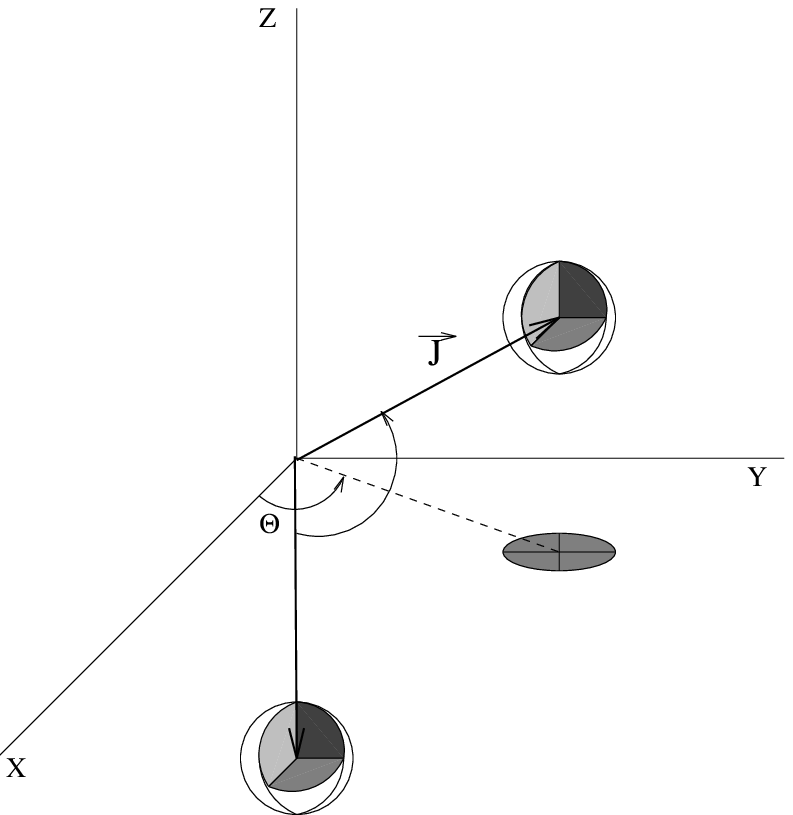}
\includegraphics{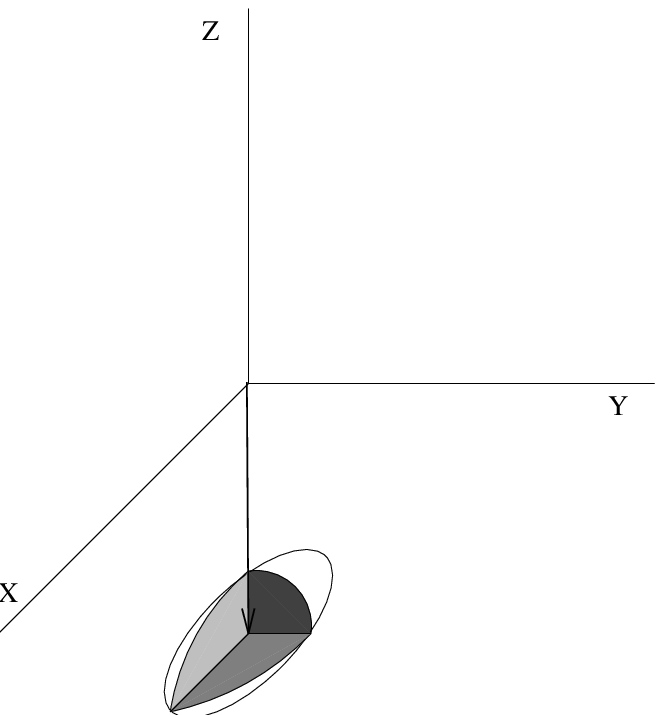}
\caption{Schematic phase space structures of uncertainties in angular 
momentum coherent and squeezed states are compared with the harmonic 
oscillator coherent and squeezed
states referring to the uncertainties. 
The figures are not drawn according
to scale. The figures for the harmonic oscillator are in two dimensional
phase space whereas the figures for the angular momentum are in three
dimensional phase space. The figures presented are : (a) harmonic
oscillator vacuum or ground state (minimum uncertainty) with the
uncertainty circle and coherent state
obtained by displacing the ground state by a distance $\vert\alpha\vert$
towards the direction making an angle $\theta$ with the space coordinate;
(b) harmonic oscillator squeezed state of vacuum or coheremt state with
uncertainty ellipse having axes ${e^{\pm r}}\over {\sqrt 2}$; (c)
angular momentum uncertainty sphere at two positions (one at pole and 
other in a rotated direction); (d) uncertainty ellipsoid for
angular momentum squeezed state .} 
\end{figure}

\pagebreak
\vspace*{-5mm}
\begin{figure}
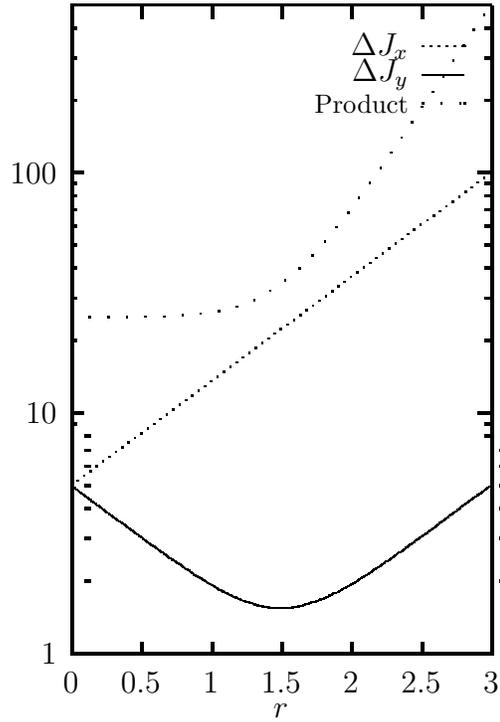

\setlength{\unitlength}{0.240900pt}
\ifx\plotpoint\undefined\newsavebox{\plotpoint}\fi
\sbox{\plotpoint}{\rule[-0.500pt]{1.000pt}{1.000pt}}%

\caption{Variation of $J_x$, $J_y$ and their product with r for single
mode squeezing with $j=50,\delta$=0 and $m=-50$ %
(Some of the marking lines, at the bottom of the figure, are shifted
towards right, which will be identified and corrected with the revised
version(s).)} 
 
\end{figure}
\pagebreak
\begin{figure}
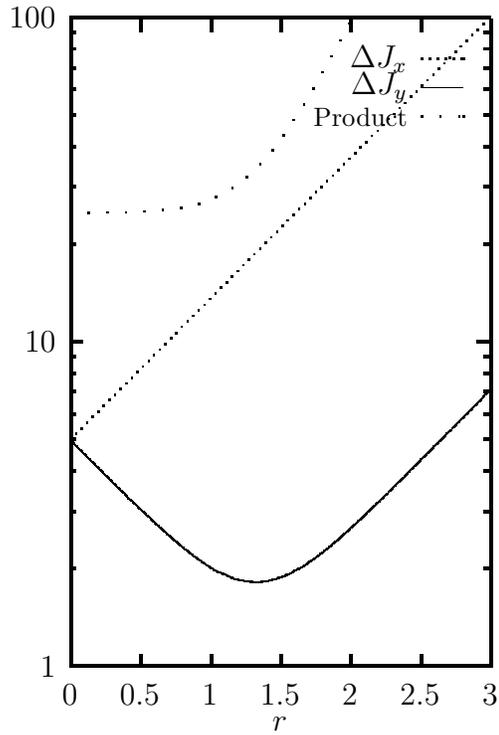

\setlength{\unitlength}{0.240900pt}
\ifx\plotpoint\undefined\newsavebox{\plotpoint}\fi
\sbox{\plotpoint}{\rule[-0.500pt]{1.000pt}{1.000pt}}%
%
\caption{Variation of $J_x$, $J_y$ and their product with r for double
mode squeezing with $j=50, m=-50, \delta$=0 and $r_\pm = r$.}
\end{figure}

\end{document}